\documentclass[conference]{IEEEtran}
\IEEEoverridecommandlockouts
\usepackage{cite}
\usepackage[utf8]{inputenc}
\usepackage{amsmath,amssymb,amsfonts}
\usepackage{algorithmic}
\usepackage{graphicx}
\usepackage{textcomp}
\usepackage{xcolor}
\usepackage{hyperref}
\def\BibTeX{{\rm B\kern-.05em{\sc i\kern-.025em b}\kern-.08em
    T\kern-.1667em\lower.7ex\hbox{E}\kern-.125emX}}

\usepackage[colorinlistoftodos]{todonotes}

\author{
\IEEEauthorblockN{Fabio N. Silva, Sergio Jimenez and George Dueñas}
\IEEEauthorblockA{Instituto Caro y Cuervo\\
Calle 10 No. 4-69, Bogotá D.C., Colombia\\
Email: [fabio.silva;sergio.jimenez;george.duenas]@caroycuervo.gov.co}}

\begin{document}

\title{Toward the Evaluation of Written Proficiency on a Collaborative Social Network for Learning Languages: Yask}

\maketitle

\begin{abstract}
Yask is an online social collaborative network for practicing languages in a framework that includes requests, answers, and votes. Since measuring linguistic competence using current approaches is difficult, expensive and in many cases imprecise, we present a new alternative approach based on social networks. Our method, called Proficiency Rank, extends the well-known Page Rank algorithm to measure the reputation of users in a collaborative social graph. First, we extended Page Rank so that it not only considers positive links (votes) but also negative links. Second, in addition to using explicit links, we also incorporate other 4 types of signals implicit in the social graph. These extensions allow Proficiency Rank to produce proficiency rankings for almost all users in the data set used, where only a minority contributes by answering, while the majority contributes only by voting. This overcomes the intrinsic limitation of Page Rank of only being able to rank the nodes that have incoming links. Our experimental validation showed that the reputation/importance of the users in Yask is significantly correlated with their language proficiency.In contrast, their written production was poorly correlated with the vocabulary profiles of the Common European Framework of Reference. In addition, we found that negative signals (votes) are considerably more informative than positive ones. We concluded that the use of this technology is a promising tool for measuring second language proficiency, even for relatively small groups of people.


\end{abstract}

\begin{IEEEkeywords}
Measuring language proficiency, Second language, Collaborative social networks
\end{IEEEkeywords}

\section{Introduction}

Quantitative evaluation of a phenomenon consists of measuring one or several variables associated to it while controlling other intervening variables that alter the measure, but that are not linked to the phenomenon, i.e. noise \cite{black1999doing}. That unavoidable situation produces differences between what is wanted to be measured and what is actually being measured. The accuracy of a particular measurement method depends significantly on whether the variables used are effectively aimed to the target and on the robustness of the method against unwanted or unavoidable factors.

The evaluation for educational purposes also obeys that principle. That is, the tools used to measure a particular skill or knowledge (for example, an exam or written test) sometimes point to a "moving target" and are usually affected by external factors. For instance, the tests for second language proficiency assessment can deviate from its intrinsic objective if they only take into account what is taught in the teaching curriculum and discard diverse cultural and linguistic backgrounds \cite{sandberg2011english}. Also they are affected by factors such as the artificial preparation (teach to the test) of the individuals being evaluated \cite{gonzalez2009alternative,menken2006teaching}, stamina to answer a long test, ability of discrimination, the handling of particular set of keywords \cite{matthiesen2017essential}, among others. Consequently, many tests of linguistic competence actually measure (noisily) many factors that may or may not be related to their actual linguistic competence. 

The current information era and the raise of social networks provide new approaches for quantitative evaluation based on the principle of the ``wisdom of the crowd'' \cite{golub2010naive}. Consider the case of StackOverflow\footnote{https://stackoverflow.com/}, a social network where computer programmers ask questions that are collaboratively answered by the on-line community. Traditionally, a programmer's degree of technical competence is determined by the use of written, oral or automated tests, which suffer from many of the aforementioned problems in the language-proficiency domain \cite{douce2005automatic}. A recent study \cite{movshovitz2013analysis} showed that the reputation gained from the social interactions on StackOverflow is an accurate predictor of the programming skills of the users of that social network.

Even more recently, a new collaborative social network for language practicing, named Yask\footnote{https://www.Yask.ai}, has been gaining popularity, recognition, and an increasing number of active users \cite{cnn_chile_2018}. Yask has a similar structure to StackOverflow, opening the research perspective of measuring the written proficiency of the users based on their interactions and votes in the social network. The methods for measuring the user importance or reputation in a social network are based on the analysis of the structure of the social graph. A well known method for that is the algorithm Page Rank \cite{page1999pagerank}. Our method, called Proficiency Rank, extends Page Rank by integrating positive, negative, implicit, and explicit signals from the social graph. In this work, we are focused on determining if Proficiency Rank is an appropriate variable for measuring the language competence of a group of users in a social network like Yask. This approach could produce a new method for language-proficiency evaluation intrinsically free from many of the issues associated with the classic approaches, while creating new challenges to overcome in the endeavor. 

\section{Background}
\subsection{Language proficiency assessment}
Language proficiency in a second foreign language (L2) comprises the ability to do ``something'' with the language but also knowing about it \cite{harsch2016proficiency}. This proficiency can be understood as pragmatic knowledge to face real-life communicative situations. After Hymes  \cite{hymes1972communicative}, two parallel approaches were suggested. The first one takes into consideration sociolinguistic and discourse abilities needed to communicate appropriately. The second one is based on the performance as a result of intertwinedness among linguistic, mental and social competence and their mutual dependence on context \cite{bachman1996language}.  However, in the educational domain there are other assumptions for proficiency \cite{cummins1979cognitive} divided mainly into two groups according to the linguistic aim, namely: the Basic Interpersonal Communication Skills (BICS) or everyday interaction skills, and the Communicative Academic Language Proficiency (CALP) or academic and  schooling knowledge communication. In terms of language testing proficiency discussion focuses on its innate nature as unitary \cite{oller1979language} or divisible \cite{palmer1981basic}. The unitary vision supposes the existence of an indivisible underlying structure \cite{tesniere1959elements}. The divisible proficiency theory \cite{palmer1981basic} holds that proficiency could be divided into subcategories e.g.: writing, speaking, listening and reading. 

Over the last two decades a ``multidimensional conceptualization of language proficiency'' has pointed out the existence of a set of different communicative skills and strategies. The Common European Framework of Reference for Languages (CEFR) \cite{council2001common}, based on a divisible language proficiency model, depicts six different ascending levels of proficiency as follows A1, A2, B1, B2, C1, and C2. The A-labels corresponds to Elementary level, B-labels to Intermediate and C-labels to Advanced. CEFR describes proficiency from overall skills and abilities to particular and less important aspects of human communication. CEFR has become a mandatory tool for teaching materials, curricula, and assessment since it was proposed in 2001. Language proficiency exams are presented into separated sections like reading, listening, speaking, writing. Each section assesses a particular skill needed for communication, expression and language understanding. The language aim for these exams is expected to cover everyday contexts.

\subsection{Automatically Assessing Writing Proficiency}

This task consist in ``predicting at which language learning stage a text can be produced or understood by an L2 learner'' \cite{pilan2016predicting} according to a scale such as the CEFR. Several linguistics features are extracted from the texts to produce quantitative proficiency predictions. Pilán et al. \cite{pilan2016predicting} proposed a method consisting of 61 linguistics features divided into five groups (length-based, lexical, morphological, syntactic and semantic features) and a Support Vector Machine (SVMs) for all their experiments of classifying essays written by L2 learners of Swedish into the CEFR levels. Their datasets were error-prone essays written by learners and error-free texts (coursebooks) written by experts, both manually labeled for CEFR levels. The best approach obtained an $F_{1}$ of $.747$ and a $\kappa^{2}$ of $.890$, which is the weighted combination of L2 coursebook texts and $60 \%$ of Swedish L2 learners’ essays. Lexical features were the most predictive measuring the proportion of tokens per CEFR level in the texts.

Tack et al. \cite{tack2017human} collected a corpus of English short answers question based on the CEFR levels and implemented an approach via a soft-voting classifier integrating a panel of five traditional models: Gaussian Naive Bayes classifier, a CART Decision Tree, a kNN classifier, a one-vs.-rest (OvR) Logistic Regressor and a OvR polynomial LibSVM Support Vector Machine. They used $695$ individual features grouped into 18 different families, among which are: lexical features, 
syntactic features,
discursive features, 
number of psycholinguistic norms. 
The best approach obtained an $F_{1}$ of $.495$ and an adjacent accuracy of $.978$. Sentence and word length, lexical features and information about the age of acquisition of words had a strong positive correlation with the assessed CEFR level. Also, the system did not have any particular difficulties in correctly predicting the lowest CEFR levels.

Yannakoudakis et al. \cite{yannakoudakis2018developing} proposed a method consisting of a corpus of $2,312$ English texts with their CEFR scores, which were assigned by a human expert, five feature types (character sequences, Parts of Speech sequences, hybrid word and Parts of Speech sequences, phrase structure rules, and errors and error rate) and a classification algorithm. The best approach obtained a Pearson $r$ of $0.7654$, a Spearman $\rho$ of $0.773$ and a $\kappa$ of $0.738$, with $0.026$ of standard error of $\kappa$. This model used the test set (consisting of 260 texts) and the PoS feature.




``My Tailor is rich!'' was a Machine learning level prediction competition in conjunction with
CAp2018\footnote{\url{http://cap2018.litislab.fr/competition-en.html}}, which task was predicting English level according to the 6 reference levels of the CEFR by analyzing written texts between 20 and 300 words and a set of characteristics calculated from these texts. The Organizing Committee provided fifty-nine feature variables, mainly shallow features based on the state of the art of stylometry and language readability. The participating systems produced predictions with very high accuracy. However, a further analysis of the results revealed that the texts contained lexical features that made the classification trivial for some systems \cite{CLEF2018}. For instance, texts labeled with C1 level were prompted by the instruction ``Write a movie review''. Therefore, the simple identification of words such as ``movie'', ``film'', ``actor'', etc., were accurate predictors of the level. They concluded that different data is needed and more research is necessary to tackle the problem.

\subsection{Collaborative Social Networks}

\begin{figure*}[ht]
\centering
\includegraphics[scale=0.50,viewport=0 70 700 580,clip]{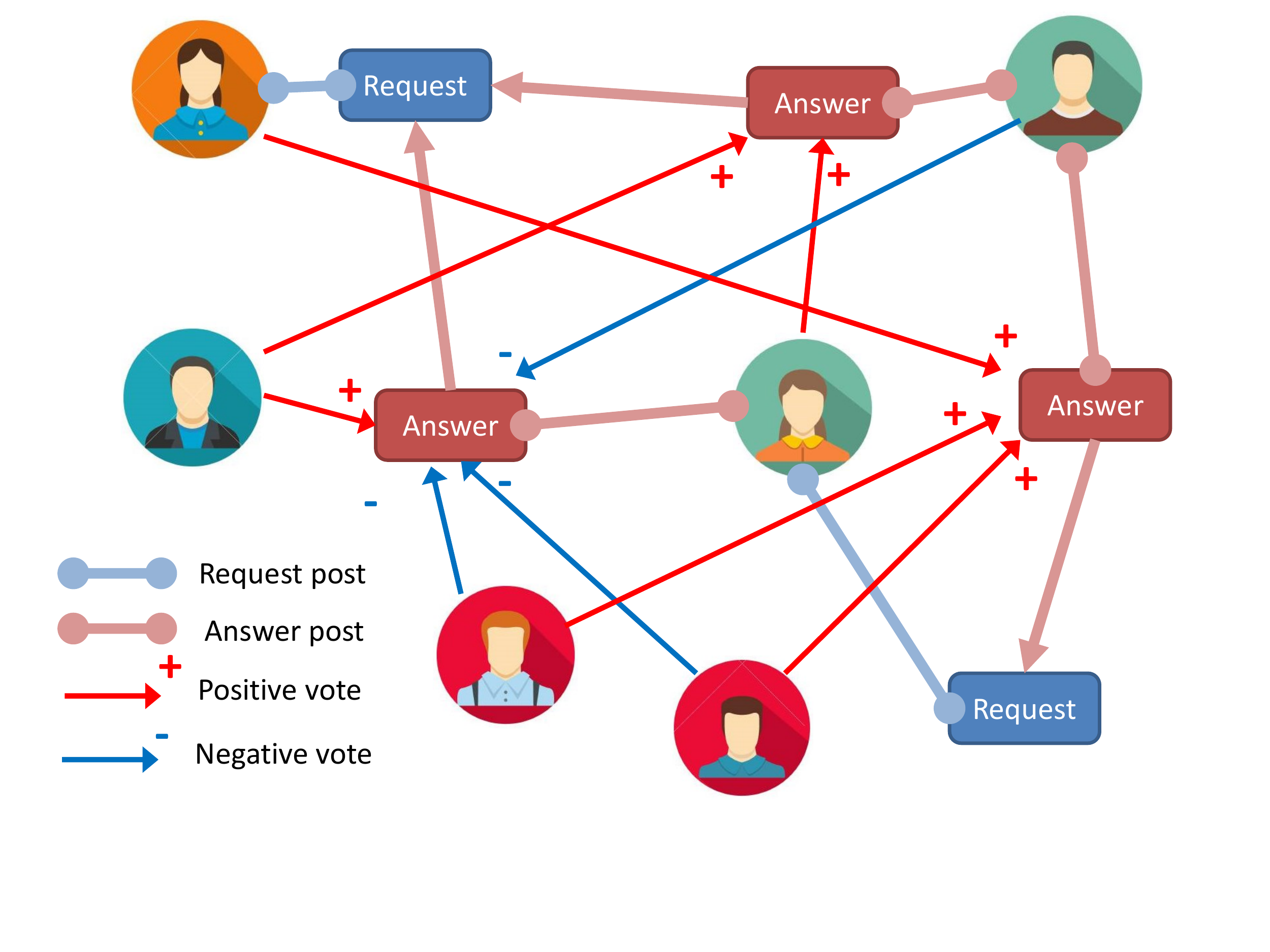}
\protect\caption{Example of a Collaborative Social Network.\label{fig:CSN}}
\end{figure*}

A collaborative social network is an arrangement of individuals with common or compatible goals that collaborate between them to achieve those goals. A particular case of this social structure is the online communities where members make requests that are addressed voluntarily by the community \cite{porter2004typology}. Members of these communities interact with each other by posting a request, answering a request or voting for in favor or against any post. Figure \ref{fig:CSN} illustrates this type social network. An example of these social networks is StackOverflow (SO), which gathers the community of computer programmers around questions and problems from that domain. There, users can make requests about computer programming, which receive voluntary answers from other users. Any answer (and also questions) receives votes from other users expressing their approval or disapproval. As a consequence of this dynamic, in the short term, the best answers are identified, and in the middle term, the skill of the users are revealed. In 2018, SO had more 9 million users and 16 million of answered questions\footnote{\url{https://stackoverflow.com/company}, \url{https://sostats.github.io/}}. Recently, SO evolved to SO Jobs, an initiative that aims to match employers and SO'users based on the reputation of the users in SO. That new approach for evaluating people skills based on social media constitute an interesting alternative to traditional methods such as exams, tests, and interviews.

\begin{figure}[ht]
\centering
\includegraphics[scale=0.17,viewport=0 100 700 1230,clip]{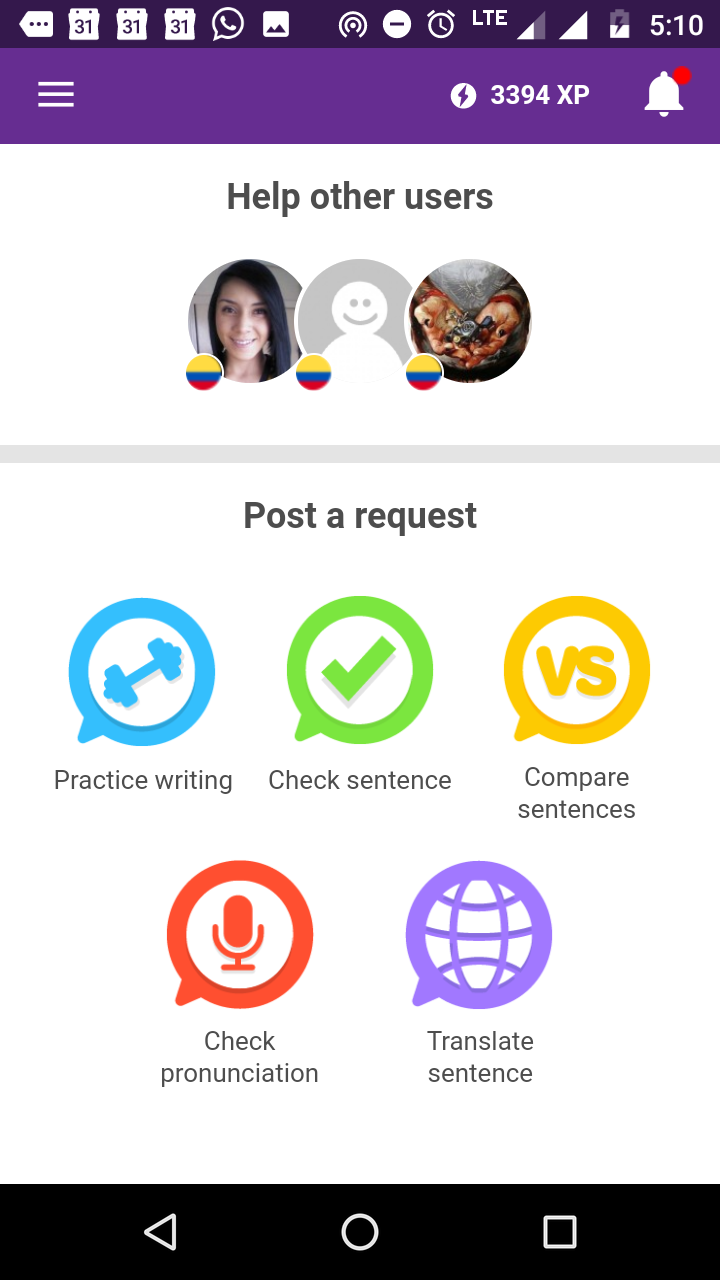}
\includegraphics[scale=0.17,viewport=0 100 700 1230,clip]{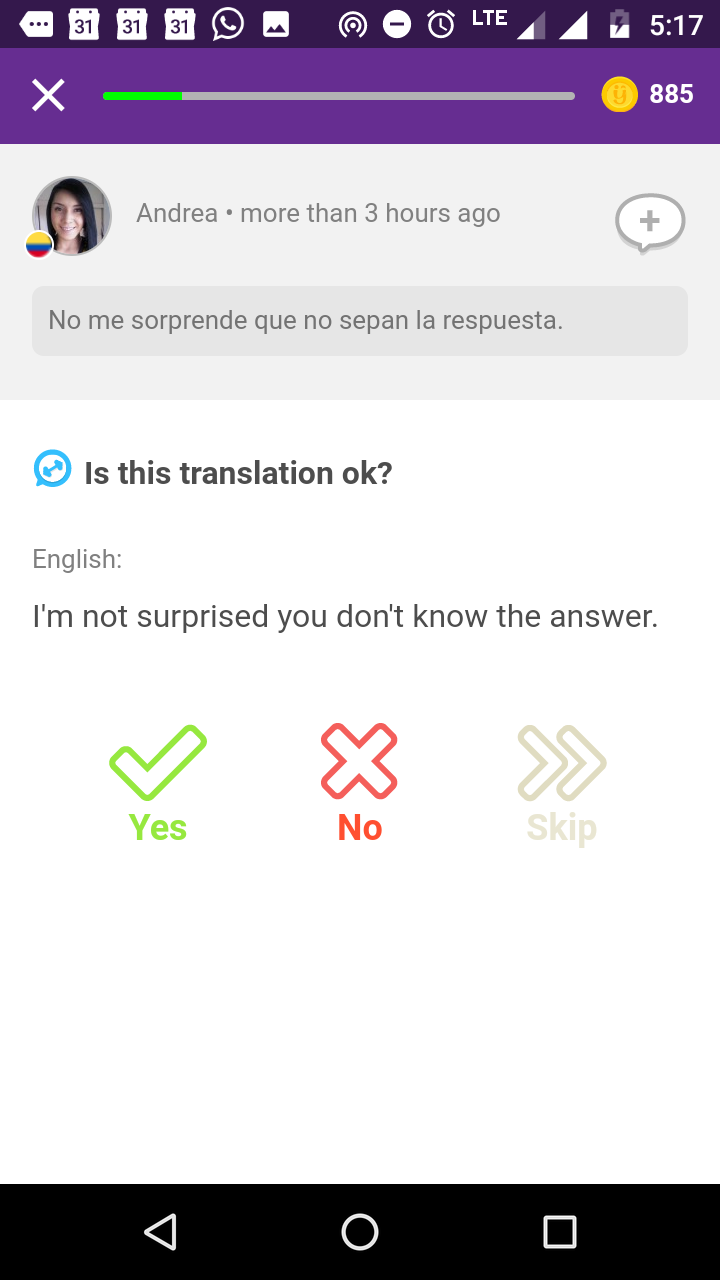}\\
\protect\caption{Screenshots from the Yask app of the main and voting screens.\label{fig:Yask}}
\end{figure}

Yask is another collaborative social network with a similar structure to SO but oriented to the learning of languages. In Yask, users contribute with their expertise in their native languages to requests made by language learners. Figure \ref{fig:Yask}a shows a screenshot of the Yask's home screen in a mobile device illustrating the type of requests supported. 
The requests are automatically sent to possible contributors, which can vote (see Figure \ref{fig:Yask}b) or answer the request by posting another answer. When a user is asked for a vote s/he has not any information about the possible previous votes of other users. When an answer reaches a certain number of votes, then it is considered solved and no more votes are requested to the community. The limit of votes per answer is not explicitly established, but answers with more than 15 votes are rare.

\subsection{Ranking Nodes' Importance in a Graph}
\begin{figure}[ht]
\centering
\includegraphics[scale=0.40,viewport=0 320 460 530,clip]{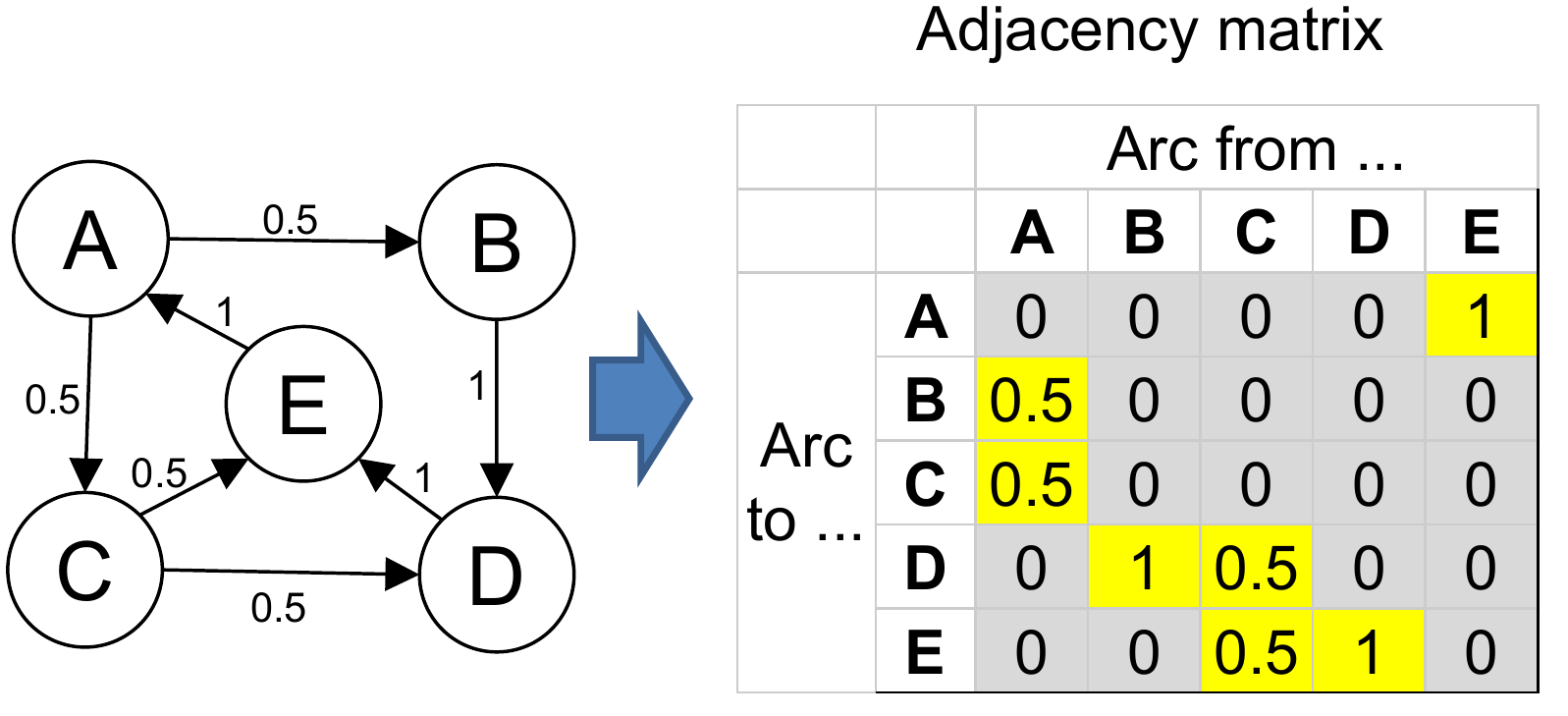}
\protect\caption{An example of a weighted graph and its adjacency matrix.\label{fig:graph}}
\end{figure}

A graph is a model consisting of a set of nodes interconnected between them by arcs or edges. The graph can be directed or undirected, where each arc has a source and destination nodes (arrows), or the arcs simply connect two nodes bidirectionally. In addition, arcs may be labeled with numeric weights, becoming a weighted graph. Figure \ref{fig:graph} illustrates a weighted-directed graph along with its representation as an adjacency matrix. Graphs are useful for modeling many problems. For instance, directed graphs can be used for modeling scientific journals (nodes) interconnected by citations (arcs) \cite{persson2010identifying,nassiri2013normalized}, the World Wide Web (web pages) interconnected by hyperlinks \cite{broder2000graph}, among others. 

In many scenarios it is necessary to determine the ``importance" or ``relevance" of the nodes in a graph by analysing solely the graph's structure \cite{han2009evaluation,han2012computing,movshovitz2013analysis}. These methods are based on the recursive hypothesis that the importance of a node depends on the importance of the nodes that have arcs towards that node. This hypothesis is equivalent to the probability of a random walker of visiting any node in the graph. Page Rank \cite{page1999pagerank} is probably the most popular method for obtaining such importance. Let $n$ be the number of nodes in a directed graph, $\mathbf{M}$ the $n\times n$ adjacency matrix, and $\mathbf{R_{t}}$ a $n\times 1$ vector of node's importance at the iteration number $t$. The Page Rank algorithm requires that the columns of $\mathbf{M}$ and $\mathbf{R_{t}}$ to be probability distributions, i.e. to sum up 1. The initial nodes' importance $\mathbf{R_{0}}$ are assigned randomly. The recursive relation of Page Rank is defined as (the operator $\cdot$ is the dot-product or matrix multiplication):

\begin{equation}
\mathbf{R_{t}} = \hat{\mathbf{M}} \cdot \mathbf{R_{t-1}}. \label{EQ:page_rank}
\end{equation}

Where the entries of the matrix $\hat{\mathbf{M}}$ are:

\begin{equation}\label{eq:damping}
\hat{m}_{i,j}=d \times m_{i,j} + \frac{(1-d)}{n}; i,j \in [1\cdots n]    
\end{equation}

Here $m_{i,j}$ are the entries of $\mathbf{M}$ and $d$ is a ``damping'' factor, which allows the random walker to escape from ``sink'' nodes, i.e. nodes only having inbound arcs. The value of $d$ is set to 0.85 for the world-wide-web problem, but it needs to be determined for other applications. After several iterations of applying eq. \ref{EQ:page_rank}, $\mathbf{R_{t} \sim R_{t-1}}$, so the algorithm converges.

By applying Page Rank to the graph showed in Figure \ref{fig:graph}, the following ranks are obtained: 0.254 for A, 0.137 for B and C, 0.207 for D, and 0.265 for E. This is the probability distribution of a random walker of visiting each node. Thus, E is the most ``popular'' node by receiving two incoming arcs. However, D, which receives also two arcs, ranks third after A, which inherits E's popularity. Finally, B and C rank last because they share the inherited popularity of A.

\section{Method: Proficiency Rank}\label{SEC:method}
Our method for Proficiency Rank consists of building two adjacency matrices, one for positive votes and another for negative. Thus, each time an answer posted by a user $A$ receives a vote form a user $B$, we draw a directed edge from the note $B$ to node $A$ either in the graph of positive votes or in  one of the negative votes. Next, we apply the Page Rank algorithm separately to each one of the graphs/matrices. Then, the two Page Rank run converge, their results are linearly combined using a parameter $\alpha$ that controls the weights of the positive and negative signals. The ranks obtained from the graph build with positive votes should increase according to language proficiency. Conversely, the rankings obtained from the graph build form negative votes are related inversely with the proficiency. Therefore, the linear combination of both rankings must be made with a negative sign. Thus, the Proficiency Rank vector $\mathbf{PR}$, containing the rank values for each user in the network, is defined as:

\begin{equation}\label{eq:proficiency_rank}
\begin{aligned}
&\mathbf{PR}=(1-\alpha)\times\mathbf{PR}_{+}-\alpha\mathbf{\times PR_{-}};\,\alpha\in\left[0,1\right];\\
&\mathbf{PR}_{+}=\mathbf{R}_{t}=\hat{\mathbf{M}}_{+}\cdot\mathbf{R}_{t-1};\\
&\mathbf{PR}_{-}=\mathbf{R}_{t}=\hat{\mathbf{M}}_{-}\cdot\mathbf{R}_{t-1}.
\end{aligned}
\end{equation}

In this definition, the first line shows the linear combination of the rankings from the positive and negative signals. The next two equations are the Page Rank runs over the positive and negative adjacency matrices, $\hat{\mathbf{M}}_{+}$ and $\hat{\mathbf{M}}_{-}$. The value of $t$ corresponds to the iteration where the Page Rank algorithm converges.

There is a limitation in the use of Page Rank, that is, that the ranking can be determined only for those users that post answers, that is, users having incoming votes. For the remaining users that only vote, the Page Ranks algorithm assigns an equal minimum ranking value. Generally, the users that only make votes outnumber significantly those who post answers making that the rankings can be obtained only for a small subset of users. To overcome this issue, we extracted ``implicit votes'' from the set of voters of a particular answer. For instance, in Figure \ref{fig:CSN}, user C and E voted contrarily the answer posted by D. Then, aside of the explicit votes of C and E toward D, it can be considered that users C and E mutually oppose producing two implicit negative votes between them. We call these votes \emph{Implicit Opposition Votes} ($iov$). We distinguish as $iov^{+}$ the implicit positive vote from C to E, and as $iov^{-}$ the implicit negative vote from E to C. Similarly, users A and C agree positively in their votes, as E and F agree negatively to the same answer. These agreements produce that we call \emph{Implicit Agreement Votes} ($iav$), which in this case produce mutual positive votes between A and C, and between E and F. We distinguish the implicit positive votes between A and C as $iav^{+}$, and those between E and F as $iav^{-}$. By considering $iov$s and $iavs$s, the number of users in the graph having incoming votes increases considerably, making possible the computation of their Proficiency Rank.

To build the matrices $\mathbf{M}_{+}$ and $\mathbf{M}_{-}$ we combine explicit and implicit votes using again a weighted linear combination:
\begin{equation}\label{eq:positive_negative_matrices}
\begin{aligned}
&\mathbf{M}_{+}=(1-\beta)\times\mathbf{M}_{exp^{+}}+\beta\times\mathbf{ M}_{iav};\,\beta\in\left[0,1\right];\\
&\mathbf{M}_{-}=(1-\delta)\times\mathbf{M}_{exp^{-}}+\delta\times\mathbf{ M}_{iov};\,\delta\in\left[0,1\right].\\
\end{aligned}
\end{equation}

Here $\mathbf{M}_{exp^{+}}$ is the adjacency matrix built using the explicit positive votes, and $\mathbf{M}_{exp^{-}}$ the equivalent with explicit negative votes. Similarly, $\mathbf{M}_{iav}$ is the adjacency matrix built using the $iav$s, and $\mathbf{M}_{iov}$ the equivalent for $iov$s. It is important to note that the entries of all the $\mathbf{M}_{*}$ matrices contain the total number of votes given by the users indexed by the columns, towards the users indexed by the rows, then the columns are normalized to sum up 1 (see Fig. \ref{fig:graph}). This differs from the traditional setting of Page Rank, where several links from a node $A$ to $B$ are treated as a single one. In addition, the matrices in eq. \ref{eq:positive_negative_matrices} need to transformed to ``hat'' versions using eq. \ref{eq:damping}.

In summary, our method has 4 parameters, namely: $d$ the damping factor of Page Rank, $\alpha$ the weighting parameter between positive and negative votes, $\beta$ the weighting parameter between explicit and implicit positive votes, and $\delta$ the analogous for negative votes. In addition, for the construction of $\mathbf{M}_{iav}$ it is necessary to determine which combination of $iav^+$ and $iav^{-}$ should be used (analogously for of $\mathbf{M}_{iov}$).

\section{Experimental Validation}
Our experiments aim to address two questions. First, to what extent the ranking methods presented in section \ref{SEC:method} are correlated with the English proficiency level of the users in Yask. Second, how much user interaction in Yask is needed to measure adequately the English proficiency level of the users.

\subsection{Dataset Description\label{sec:data}}
The data used in the experiments was extracted manually using the Yask application for smartphones. Firstly, we created a user and interacted actively during approximately one month by posting answers to other users requests in English. Next, we recorded all the users and votes to all of our answers. Finally, we selected 10 users with Spanish at native level and English between beginner to advanced level and recorded all posts, users and votes related to their requests. The result was a graph with 377 users (nodes), 1,571 positive votes (arcs), and 490 negative votes (nodes). The English proficiency level of the users, as established by the users, is distributed as: 69 `Native', 52 `Fluid', 66 `Advanced', 140 `Intermediate', and 50 `Beginner'. The number of requests asked by the users is 179, while the number of answers to those requests is 412. These 412 answers were made by only 107 users, which are the only ones able to receive votes. Approximately, the 50\% of the answers were posted by 10 users, among them are Google Translate and the Yask Bot, which is an automatic response of a previously answered request in Yask. This observation confirmed our assumption that users that contribute with answers are a minority in comparison with those with contribute with votes. This process of extraction was performed during January 2019.

\subsection{Experimental Setup}
The gold standard for comparing the rankings produced by the methods is the English level that the users manifested freely when they signed up into Yask, which we assume to be true. We replaced the categorical levels by a simple numerical scale as follows: 5 for `Native', 4 for `Fluid', 3 for `Advanced', 2 for `Intermediate', and 1 for `Beginner'. The evaluation measure to compare the degree of agreement between the gold standard and the produced rankings is the Spearman's rank correlation. 

Each particular configuration of Proficiency Rank consists of a selection of types of positive and negative votes. Positive votes can be a combination of a selection from $\mathbf{M}_{exp^{+}}$ ,$\mathbf{M}_{iav^{+}}$, and $\mathbf{M}_{iav^{-}}$. Similarly, negative votes come from $\mathbf{M}_{exp^{-}}$ ,$\mathbf{M}_{iov^{+}}$, and $\mathbf{M}_{iov^{-}}$. Once the types of votes to use have been established for a configuration, the parameters $[d,\alpha,\beta,\delta]$ are determined in a search grid with a resolution of 0.1. Then, the grid resolution was increased to 0.05 in the vicinity of the current best configuration, and again the resolution is increased until 0.01. The function to optimize is the average of Spearman's $r$ correlations between Proficiency Rank and the gold standard for different subsets of the users filtered by a threshold of $\theta$ representing the minimum number of incoming votes. $\theta$ is incremented from 1 to the maximum number of votes obtained by the top-voted user. Clearly, as $\theta$ increases the number of users that surpass that threshold reduces. For the average calculation, we considered only significant correlations with $p<0.01$. 

Table \ref{tab:exp_setup} shows seven possible configurations that we consider interesting to discuss. The last row reports the total number of votes on each category of the type of votes found in the data.  
The baseline method consists of the total number of incoming votes per user for each configuration. This measure quantifies the possible undesired effect that the amount of activity of the users in the network being correlated with their language proficiency. The results for all the 377 users for each configuration can be seen in the last column in Table \ref{tab:exp_setup}.

\begin{table*}[t]
\centering 
\protect\caption{The seven configurations of Proficiency Rank used in the experiments with their optimal set of parameters.\label{tab:exp_setup}}
\setlength\tabcolsep{2.7pt} 
\begin{tabular}{cccccccccccrr}
\multicolumn{1}{l}{} & \multicolumn{1}{l}{} & \multicolumn{4}{c}{Positive Votes} & \multicolumn{1}{l}{} &
\multicolumn{4}{c}{Negative Votes} \\
\cline{3-6} \cline{8-11}
\multicolumn{1}{l}{Conf.}&$d$&$exp^{+}$&$\beta$&$iav^{+}$&$iav^{-}$&$\alpha$&$exp^{-}$&$\delta$&$iov^{+}$&$iov^{-}$&Votes&Baseline\\
\hline
conf1&0.86&\checkmark&0.00&&&0.79&\checkmark&0.00&&&2,061&0.186*\\
conf2&0.80&\checkmark&0.90&&\checkmark&0.78&\checkmark&0.40&&\checkmark&3,873&0.011\\
conf3&0.85&\checkmark&0.00&&&0.85&\checkmark&0.15&\checkmark&\checkmark&3,393&-0.085\\
conf4&0.98&\checkmark&0.40&\checkmark&&0.39&\checkmark&0.74&\checkmark&&10,125&-0.147\\
conf5&0.90&\checkmark&0.10&\checkmark&\checkmark&0.65&\checkmark&0.00&&&10,605&-0.094\\
conf6&0.85&\checkmark&0.14&\checkmark&\checkmark&0.66&\checkmark&0.15 &\checkmark&\checkmark&11,937&-0.126*\\
conf7&0.89&\checkmark&0.53&&\checkmark&0.85&\checkmark&0.20&\checkmark&\checkmark&4,539&0.005\\
\hline
\multicolumn{2}{l}{Num. of votes:}&1,571&&7,398&1,146&&490&&666&666&11,937\\
\end{tabular}\\
* significant $p<=0.05$.
\end{table*}

\subsection{CEFR Baseline}
We provide an additional test bed for comparison that reflects the methods of the current language teaching curricula. For that, we used the English Vocabulary Profiles for the Common European Framework of Reference (CEFR) provided by \emph{EnglishProfile}\footnote{\url{https://englishprofile.org/wordlists}}, a non-profit organization devoted to produce resources for teaching English aligned with the CEFR levels (i.e. A1, A2, to C2). They provide manually curated word lists obtained from the Cambridge Learner Corpus \cite{nicholls2003cambridge}, which represent the vocabulary profile for each CEFR level\footnote{we use the lists compiled at \url{https://www.toe.gr/course/view.php?id=27}}. Table \ref{tab:CEFR_profiles} shows the number of words on  the vocabulary profiles of each level (table's diagonal) and the number of common words between levels.\\

We used the texts of the answers written by the Yask's users in combination with the CEFR vocabulary profiles to determine the level for each user. For that, we obtained all $w_{u,l}$, which is the number of common words between the set of words derived from the answers written by user $u$ and the vocabulary profile corresponding to the level $l$. Since, the CEFR levels are meant to correspond to a linear progression of proficiency in English, we assigned increasing weights to each level. Thus, the proficiency level $P_{u}$ for a particular user $u$ is computed with a weighted average as follows:
\begin{equation}\label{eq:CEFR_baseline}
P_{u}=\frac{\sum_{i=1}^{6}i\cdot w_{u,l_{i}}}{\sum_{i=1}^{6} w_{u,l_{i}}}
\end{equation}
Where $l_{1}$ corresponds to the level A1, $l_{2}$ to A2, until $l_{6}$ to C2. Therefore, $P_{u}$ is a number between 1 and 6...\\

To evaluate the soundness of this baseline, we applied eq. \ref{eq:CEFR_baseline} to the 27,306 texts from the CAp2018 training dataset\footnote{\url{http://cap2018.litislab.fr/competition-en.html}}. The obtained values were compared against the gold standard levels in the same dataset. We observed a Spearman'$r$ rank correlation of 0.65 with $p<0.0001$. Clearly, the proposed baseline represents the CEFR levels of English proficiency.

\begin{table}[]
    \centering
    \caption{Number of common words between the English Vocabulary Profiles obtained from the Cambridge Learner Corpus for the CEFR levels.}
    \label{tab:CEFR_profiles}
    \begin{tabular}{c|cccccc}
         &A1&A2&B1&B2&C1&C2  \\
         \hline
         A1&541&217&214&188&156&197 \\
         A2&-&1038&415&385&262&348 \\
         B1&-&-&1806&731&394&533 \\
         B2&-&-&-&2495&536&717 \\
         C1&-&-&-&-&1701&575 \\
         C2&-&-&-&-&-&2182 \\
    \end{tabular}
\end{table}

\subsection{Results}

\begin{figure}
\centering 
\includegraphics[scale=0.48, viewport=40 230 600 550,clip]{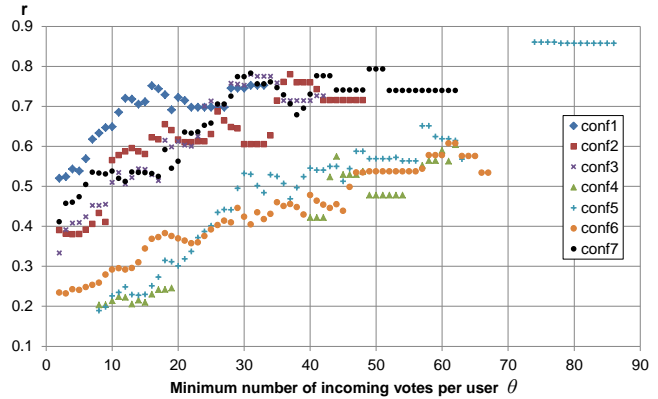}
\protect\caption{Results of the tested Proficiency Rank configurations for different sets of users having at least $\theta$ incoming votes.\label{fig:pr_votes}}
\end{figure}

\begin{figure}
\centering 
\includegraphics[scale=0.48,viewport=40 240 600 550,clip]{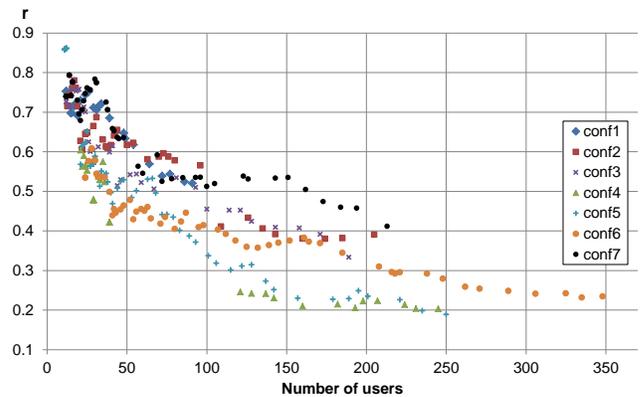}
\protect\caption{Results of the tested Proficiency Rank configurations for different sizes of sets of users. \label{fig:pr_users}}
\end{figure}

Figure \ref{fig:pr_votes} shows the results obtained by the Proficiency Rank configurations from \emph{conf1} to \emph{conf7}. The vertical axis corresponds to the Spearman's rank correlation $r$ between the Proficiency Rank produced by each configuration versus the gold standard measured in a subset of the users filtered by $\theta$ (horizontal axis). In figure \ref{fig:pr_votes}, all seven configurations increase as $\theta$ increases. The total number of votes considered for applying the threshold $\theta$ varies on each configuration. For instance, configuration \emph{conf1} considered only 2,061 votes (1,571+490), and \emph{conf6} used all the 11,937 available explicit and implicit votes. As $\theta$ increases, the number of users that fulfill that threshold decreases. Figure \ref{fig:pr_users} shows the same results but replacing the abscissa $\theta$ by the number of users, producing a decreasing tendency for all configurations. In general, the best configurations are those that shape the upper-bound in both figures. Note that all configurations outperformed their corresponding baselines by a wide margin.

\subsection{Reproducibility of Experiments}
To reproduce the experiments of this study it is necessary to ask Yaks's representatives for the data used in this study or the consent for extracting a data sample from Yask with a methodology similar to the one described in subsection \ref{sec:data}. The code written for Python 3.7 for reproducing the experiments and a sample with the data format is available at \url{https:\\}

\begin{figure}
\centering 
\includegraphics[scale=0.48,viewport=40 200 600 570,clip]{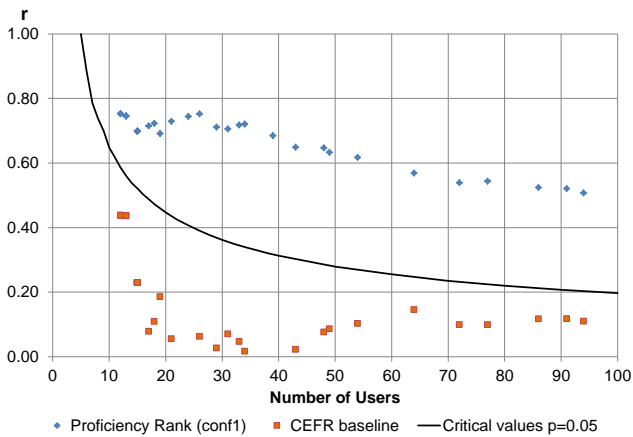}
\protect\caption{Results of Proficiency Rank \emph{conf1} in comparison with the CEFR baseline.\label{fig:CEFR_baseline}}
\end{figure}

Figure \ref{fig:CEFR_baseline} shows a comparison of the results obtained by Proficiency Rank and the CEFR baseline proposed in eq. \ref{eq:CEFR_baseline}. This comparison is only possible against \emph{conf1} because the users who received votes are the only ones who wrote answers. Thus, the $P$ baseline for each user is computed by aggregating all the answers written by each user. In addition, this figure includes a line with the critical values for the Spearman's $r$ rank coefficient for $p=0.05$.

\subsection{Discussion}
 Let us discuss the results obtained by \emph{conf1}, a configuration composed only explicit votes, therefore the one with least number of votes. \emph{Conf1} achieved the best results when $\theta$ varies between 1 and 25. This result indicates that explicit votes are the strongest signal in the social graph. However, explicit votes are in short supply producing only significant Proficiency Ranks for only a maximum of 91 users out of 379. Figure \ref{fig:pr_users} shows that other configurations produce Proficiency Ranks for far more users. The optimal value for the parameter $d$, when using \emph{conf1}, coincides with the default value for the equivalent damping parameter for Page Rank ($d=0.85$) \cite{page1999pagerank}. Parameter $\alpha=0.79$ indicates that even though, the number of explicit negative votes is relatively small ($|exp^{-}|=490$), they weight much more than the explicit positive votes ($|exp^{+}|=1571$). In addition to the showed experiments, we tested the results of Page Rank for positive and negative votes separately. However, no significant correlation was observed ($p<0.01$) for any possible value of $\theta$. In the first test (only votes in  $exp^{+}$), the number of explicit positive votes seemed to be sufficient for 377 users, but their lack of informativeness could explain the poor results. Contrarily, the explicit negative votes are highly informative, but only 490 edges for 377 nodes produces a very sparse graph. This result proves that our method for combining the positive and negative votes using a linear combination controlled by $\alpha$ is effective in comparison with Page Rank using alternatively positive and negative votes.
 
 Figure \ref{fig:pr_votes} shows that configurations \emph{conf1}, \emph{conf2}, \emph{conf3}, and \emph{conf7} outperformed the others. These configurations have in common the use or $iov^{-}$ and the disregard of $iov^{+}$. Again this result shows the preponderance of a few negative signals versus a large number of positive signals. We consider that the best configuration is \emph{conf7}. Figure \ref{fig:pr_users} shows that \emph{conf7} performs among the best configurations in the range from 10 to 100 users, and it is the best for more than 100 users. Although, \emph{conf5} and \emph{conf6} produce Proficieny Ranks for more than 200 users, their correlations are considerably lower than those of \emph{conf7}.
 
The top result ($r=0.86$) was obtained using \emph{conf5} and $\theta>75$. That result corresponds to a set of 12 users having each one more than 75 incoming votes. In spite of being a small subset of users, the observed correlation was highly significant, $p=0.000323$. This result is somehow unexpected given the poor performance of \emph{conf5} for larger sets of users. However, this result suggests that high correlations can be achieved if there is enough information (votes) associated with each user. To provide conclusive proof of that point it would be necessary to carry out experiments with a considerably larger dataset. 

Regarding baseline results (see the last column in Table \ref{tab:exp_setup}, it is clear that the effect of the amount of user activity in the collaborative social network is poorly correlated with language proficiency. Only configurations \emph{conf1} and \emph{conf6} obtained significant correlations ($p-value<0.05$) but with a considerable margin to the lowest Proficiency Rank results. This result indicates that the measure of proficiency is mostly independent of the amount of activity of the users. 

The results of the comparison between Proficiency Rank and the CEFR baseline show that there is an possible mismatch between the targets that each measure aims. Figure \ref{fig:CEFR_baseline} shows that for all subsets of users (controlled by $\theta$) Proficiency Rank produces significant correlations, while the CEFR baseline does not. The fact that the CEFR baseline is strongly correlated with a large corpus produced by learners in a curriculum based on CEFR, but poorly correlated with our gold standard, provides empirical evidence of the misalignment of that curriculum with the language proficiency perceived by the Yask's users. It means that there is only a loose relation between the CEFR vocabulary profiles and the self-perceived proficiency in the written modality. Therefore, it seems that native speakers are not represented by these profiles, nor are the beginners or any other intermediate level.  However, the validity of the current curricula and their standardized tests has been long discussed and accepted by the academia and the language teaching and assessment industry \cite{chapelle2011toefl}. 
Although, some academics criticize current approaches \cite{alderson2007cefr}, our results seems to be the first empirical evidence of the disagreement between the ``wisdom of the crowd'' and the written language proficiency tests based on current curricula. This result confirms the difficulty on the construction of valid language assessment tests and the potential of the Social Computing technologies in that area.

\subsection{Research Perspectives} 
In general, it is possible to say that the results provided by Proficiency Rank are good predictors of the self-evaluations of the users. That is, the users ordered by Proficiency Rank are arranged from ``Native'' to ``Beginner'' level meaningfully. It is important to note that the method is independent of the English language and it is not related to any learning curriculum. However, the construction of an assessment tool based on this discovery requires more research. For instance, in our experiments, the Yask users didn't expect to be evaluated, therefore fraud or artificial preparation are not considered issues. The control of these issues in proficiency evaluation based on social interaction is an interesting research perspective. Similarly, the degree of complexity and difficulty of the requests is Yask is distributed accordingly with the proficiency of the users, which is roughly uniform. In an evaluation scenario, the requests should be provided by an evaluation authority and their difficulty should be controlled. Clearly, extremely difficult or trivial requests can hinder the overall evaluation of the users. In addition, those artificial requests should promote divided voting pools to produce enough negative votes for Proficiency Rank. The determination of an appropriate set of initial requests is also an interesting research topic. 

It is also important to note that our method is independent of the modality if the requests. In our experiments, we used only the written modality, but that feature has not been used by Proficiency Rank. Therefore, the requests could include any type of media opening the perspective of constructing requests based on listening, pronunciation, conversation, translation, etc. The use of these modalities in Proficiency Rank is another potential research direction.




\section{Conclusions}

We presented Proficiency Rank, a method for measuring user importance in a collaborative social network. By testing Proficiency Rank in a sample of the Yask community, we observed that the rankings obtained by our method are highly correlated with the language proficiency of the users. We carried out experiments that revealed how different amount of users interactions (postings and votes) produce different intensities of that correlation. In addition, we observed that the most informative signal in a collaborative social network is the negative votes. However, the best configurations of Proficiency Rank were obtained by linear combinations of positive, negative, explicit and implicit votes.

\bibliographystyle{plain}

\end{document}